# The effects of different quantum feedback operator types on the parameter precision of detection efficiency in optimal quantum estimation

Shao-Qiang Ma (马少强)[1], Han-Jie Zhu(朱汉杰)[1], Guofeng Zhang(张国锋)[1,2,3,*]

[1]*Key Laboratory of Micro-Nano Measurement-Manipulation and Physics (Ministry of Education), School of Physics and Nuclear Energy Engineering; State Key Laboratory of Software Development Environment, Beihang University, Xueyuan Road No. 37, Beijing 100191, China*

[2]*State Key Laboratory of Low-Dimensional Quantum Physics, Tsinghua University, Beijing 100084, China*

[3]*Key Laboratory of Quantum Information, University of Science and Technology of China, Chinese Academy of Sciences, Hefei 230026, China*

**Abstract:** The effects of different quantum feedback types on the estimation precision of the detection efficiency are studied. It is found that the precision can be more effective enhanced by a certain feedback type through comparing these feedbacks and the precision has a positive relation with detection efficiency for the optimal feedback when the system reach the state of dynamic balance. In addition, the bigger the proportion of $|1\rangle$ is the higher the precision is and we will not obtain any information about the parameter to be estimated if $|0\rangle$ is chosen as initial state for the feedback type $\lambda \sigma_z$.

## I. INTRODUCTION

Quantum metrology[1], a science researching on the limit of the precision of the parameter to be estimated determined by quantum mechanics, which is widely used in improving time and frequency standards[2-4], detecting gravitational waves[5], and so on, has excited wide enthusiasm in recent years. Quantum Fisher information (QFI) plays a vital role in quantum metrology and quantum estimation theory. According to quantum Cramer-Rao inequality, a larger QFI implies that the parameter can be estimated with a higher precision. And therefore one of the basic ideas about quantum metrology is to improve QFI and beyond the shot-noise limit by making use of nonclassical resources, such as maximally path-entangled NOON state, entangled state or quantum feedback. Quantum feedback [6-9], which manipulates the system based on the

---

information acquired by measurement of the controlled system, has been considered as one of the essential ways to suppress decoherence and improves the precision of parameter estimation [10,11].

In addition, as we all know the detection efficiency is a crucial physical quantity to judge the detector quality, and thus the improvement of the estimation accuracy of the detection efficiency become more and more important. Due to the mechanism of quantum feedback, we can deduce that the system can be controlled effectively only when the feedback can vary with the information we obtained. However, we find no system can offer the feedback of this type to improve QFI about the detection efficiency of the detector. In order to realize the effective control of the system and the improvement of QFI about the detection efficiency by real-time feedback, we should first research the effects of the different types of feedback on the QFI. Therefore, the main purpose of the paper is to investigate the use of the different quantum feedback types to enhance the QFI about the detection efficiency of the detector. As the most commonly used to build blocks for quantum information processing, qubit systems have played an irreplaceable role not only in theoretical analysis but also in experimental tests due to its unique properties. What's more, inspired by recent experimental achievement on quantum feedback control on the qubit system, we will take the dissipative cavity which provides the quantum feedback to the qubit as the platform to carry out our research.

This paper is organized as follows. The basic properties of the QFI are reviewed in Sec. II. And then in Sec. III, the physical model, a qubit interacting with a dissipative cavity which provides the feedback to the qubit, is introduced. In Sec. IV we find that the QFI is improved by feedback and discuss the effect of different types of feedback on the QFI. Finally, a summary is provided in the last section.

## II. QUANTUM FISHER INFORMATION

In this subsection, the basic properties of QFI will be reviewed. The single parameter to be estimated is denoted by $\varphi$, and $p_i(\varphi)$ represent the probability density with measurement outcome $\{x_i\}$ for a discrete observable X conditioned on the fixed parameter $\varphi$. The classical Cramer-Rao inequality gives the bound of the variance $\text{Var}(\hat{\varphi})$ for an unbiased estimator $\hat{\varphi}$,

$$\text{Var}(\hat{\varphi}) \geq \frac{1}{H_\varphi}, \tag{1}$$

where the classical Fisher information is defined as $H_\varphi = \sum_i p_i(\varphi)[\partial \ln p_i(\varphi)/\partial \varphi]^2$ [12].

In order to extend to quantum mechanics, the symmetric logarithmic derivative $L_\varphi$ determined by $\partial \rho_\varphi / \partial \varphi = (\rho_\varphi L_\varphi + L_\varphi \rho_\varphi)/2$ should be introduced. The so called quantum Cramer-Rao inequality provides a lower bound for variance of any unbiased estimator [12-14]:

$$\text{Var}(\hat{\varphi}) \geq \frac{1}{H_\varphi} \geq \frac{1}{F_\varphi}, \qquad (2)$$

in which the QFI of a quantum state $\rho_\varphi$ with respect to the parameter $\varphi$ is defined as[15]:

$$F_\varphi = \text{Tr}(\rho_\varphi L_\varphi^2). \qquad (3)$$

As to the achievability of the lower bound in the Eq. (2) [16-20], there are two conditions to be met. First we should select the optimal unbiased estimator and the Maximum likelihood estimator (MLE) advocated by Nagaoka is considered to be the optimal one [17, 21]. And then we ought to choose the best positive-operator-valued-measure (POVM) to satisfy the condition (B1-B4) of the reference [18]. In addition, taking advantage of the decomposition $\rho_\varphi = \sum_k \lambda_k |k\rangle\langle k|$, one can obtain the concrete form of QFI

$$F_\varphi = \sum_{k,\lambda_k>0} \frac{(\partial_\varphi \lambda_k)^2}{\lambda_k} + \sum_{k,k'\lambda_k+\lambda_{k'}>0} \frac{2(\lambda_k - \lambda_{k'})^2}{\lambda_k + \lambda_{k'}} |\langle k|\partial_\varphi k'\rangle| . \qquad (4)$$

The first term in this equation is the classical Fisher information of the probability distribution corresponding to the eigenvalues of the density operator, and the second term can be considered as the quantum contribution. And the symmetry logarithmic derivative takes the following form:

$$L_\varphi = \sum_{k,k'\lambda_k+\lambda_{k'}>0} \frac{2\langle k|\partial_\varphi \rho_\varphi|k'\rangle}{\lambda_k + \lambda_{k'}} |k\rangle\langle k'|. \qquad (5)$$

For the general quantum system, the form of $F_\varphi$ is complicated. Fortunately, the QFI of the two-dimensional density matrix has been obtained explicitly as [22,23]

$$F_\varphi = \text{Tr}\left[(\partial_\varphi \rho)^2\right] + \frac{1}{\text{Det}(\rho)} \text{Tr}\left[(\rho\, \partial_\varphi \rho)^2\right] \qquad (6)$$

## III. THE PHYSICAL MODEL

The dynamics of a single atom resonantly coupled to a single-mode cavity, which is driven by a laser field with Rabi frequency $\Omega$ and damped with decay rate $\kappa$, will be introduced. The model is shown in Fig. 1.

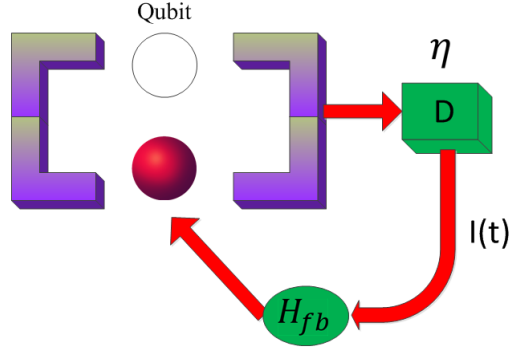

Fig. 1: Schematic view of the model, the qubit is coupled to a heavily damped cavity. The feedback Hamiltonian is applied to the atoms according to the homodyne current I(t) derived from detector D, the detection efficiency is denoted by $\eta$.

The coupling strength between the atom and the cavity is $g$, the two levels of the qubit are $|0\rangle$ and $|1\rangle$ and the spontaneous decay of the atom is $\gamma$. When the cavity mode is adiabatically eliminated under the condition that the cavity decay $\kappa$ is much larger than the other relevant frequencies of the system, an effective damping rate $\mathcal{T} = g^2/\kappa$ can be acquired [24]. In the limit $\mathcal{T} \gg \gamma$, the spontaneous emission of the atoms is neglected. Then the dynamical evolution of this system can be described by the Dick model [24-27]:

$$\frac{d\rho}{dt} = -i[\,\Omega\sigma_x, \rho\,] + D(\mathcal{T}\sigma_-)\rho, \qquad (7)$$

where $\rho$ is the density matrix of the qubit, $\Omega\sigma_x$ represents the driving of the laser, the Pauli operators $\sigma_- = |0\rangle\langle 1|$ $\sigma_+ = |1\rangle\langle 0|$ is the lowering and raising operators of the qubit respectively and $D(\mathcal{T}\sigma_-)\rho = (\mathcal{T}\sigma_-)\rho(\mathcal{T}\sigma_-)^+ - [(\mathcal{T}\sigma_-)^+(\mathcal{T}\sigma_-)\rho + \rho(\mathcal{T}\sigma_-)^+(\mathcal{T}\sigma_-)]/2$ represents the irreversible evolution induced by the interaction between the system and the environment with the jump operator $\mathcal{T}\sigma_-$. we will take $\mathcal{T} = 1$ in the following to simplify the calculation.

Now let us consider the system subject to the feedback, as shown in Fig. 1: The output from the cavity is measured by a detector D, and then the signal $I(t)$ from the detector D triggers a time-continuous feedback Hamiltonian. In the homodyne-based scheme, the detector registers a continuous photocurrent and the feedback Hamiltonian is constantly applied to the system [21]. The master equation becomes [11, 21, 26-28] (for more detail please see Appendix)

$$\frac{d\rho}{dt} = -i\left[\Omega\sigma_x + \frac{1}{2}(\sigma_+ F + F\sigma_-), \rho\right] + D(\sigma_- - iF)\rho, \qquad (8)$$

in which $F$ is the feedback Hamiltonian. Practically, the detection efficiency of the detector, denoted by $\eta$, is less than 1. The modified master equation takes the form [9]

$$\frac{d\rho}{dt} = -i\left[\Omega\sigma_x + \frac{1}{2}(\sigma_+ F + F\sigma_-), \rho\right] + D(\sigma_- - iF)\rho + D\left(\sqrt{\frac{1-\eta}{\eta}}F\right)\rho. \qquad (9)$$

In this paper, we will not take the driving of the laser into consideration, namely $\Omega=0$.

## IV. THE EFFECT OF DIFFERENT FEEDBACK ON QFI

In the following, we adopt the quantum parameter estimation theory to estimate the efficiency of the detection so as to investigate the properties of detector. The arbitrary Hermitian operator of qubit systems can be denoted by $\varepsilon_1 I + \varepsilon_2 \vec{a}\cdot\vec{\sigma}$, where $\{\varepsilon_1, \varepsilon_2\}$ are real parameters, $\vec{a} \in R_3$ are unit vectors, and $\vec{\sigma} = (\sigma_x\ \sigma_y\ \sigma_z)$ are standard Pauli matrices. Therefore, in order not to lose generality, three different Hermitian operator $I$, $\lambda(\sin[\beta]\sigma_x + \cos[\beta]\sigma_y)$ and $\lambda\sigma_z$ are selected as the feedback operator to study the effects of different feedback types on QFI of the detection efficiency.

### A. The identical feedback

In order to compare with the following two types of feedback, we first investigate the identity feedback. As we all know the identity operator has no effect on the system, therefore one can hold the view that the case choosing identity operator equal to the case choosing no feedback. After a simple calculation, one can find that the QFI equal to zero for the identity feedback. In fact, the result can be easily explained by the physical meaning of identical feedback. As we all know the efficiency of the detection is a parameter from the measurement, and thus the system cannot obtain any information about the efficiency parameter when we choose the identical feedback which carries information about the measurement but has no effect on the system.

### B. The feedback $\lambda(\sin[\beta]\sigma_x + \cos[\beta]\sigma_y)$

In the following we will study the feedback types $\lambda(\sin[\beta]\sigma_x + \cos[\beta])\sigma_y)$. For a superposition initial state $|\varphi_0\rangle = \cos[\alpha]|0\rangle + \sin[\alpha]|1\rangle$, we first concentrate on $\alpha = \pi/2$. The evolved density matrix can be exactly solved by substituting $F = \lambda(\sin[\beta]\sigma_x + \cos[\beta])\sigma_y)$ and $|\varphi_0\rangle = |1\rangle$ into Eq. (9), which is given as:

$$\rho(t) = \begin{pmatrix} \rho_{11}(t) & \rho_{12}(t) \\ \rho_{12}^*(t) & 1 - \rho_{11}(t) \end{pmatrix}, \qquad (10)$$

with the elements

$$\rho_{11}(t) = \frac{(\eta + (1+\chi)\lambda^2 + 2\eta\lambda\cos[\beta])}{\chi(\eta + 2\lambda^2 + 2\eta\lambda\cos[\beta])}, \qquad (11)$$

$$\rho_{12}(t) = \rho_{12}^*(t) = 0, \qquad (12)$$

in which, $\chi = \exp(t(\eta + 2\lambda^2 + 2\eta\lambda\cos[\beta])/\eta)$. Adopting Eqs. (6) and (10) by some straightforward calculations, the QFI about the detection efficiency is obtained as

$$F_\eta(t) = \frac{\lambda^4(\eta^2(1-\chi+2t)+6\eta t\lambda^2+4t\lambda^4+\Upsilon\Lambda)^2}{(-1+\chi)\eta^4\Theta(\Theta-\lambda^2)^2(\eta+(1+\chi)\lambda^2+\Upsilon)}, \quad (13)$$

in which $\Upsilon = 2\eta\lambda\cos[\beta]$, $\Lambda = \eta - \chi\eta + 4\eta t + 6t\lambda^2 + 2t\Upsilon$ and $\Theta = \eta + \lambda^2 + \Upsilon$. Time evolution of $F_\eta(t)$ for different quantum feedback types is shown in Fig.2.

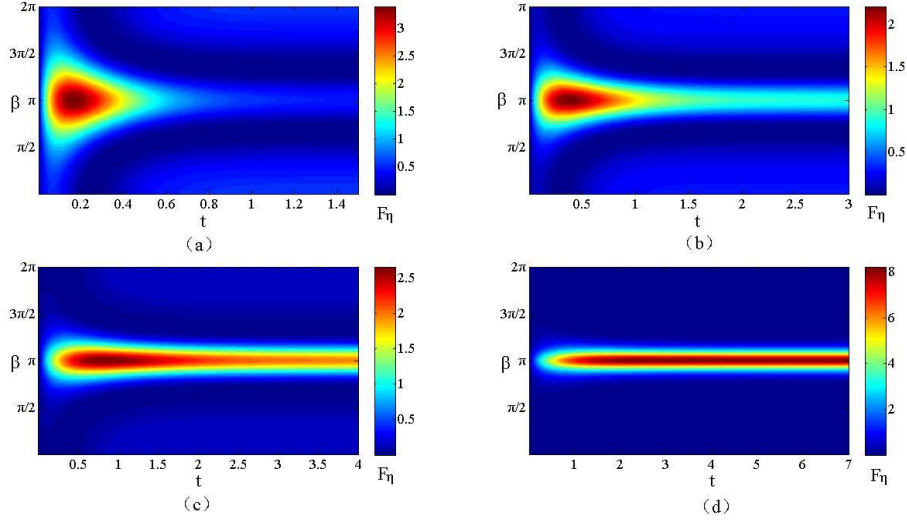

Fig.2: Time evolution of the QFI for the different value of $\eta$, here choose $|1\rangle$ as initial state, and take $\eta = 0.3$ in (a), $\eta = 0.5$ in (b), $\eta = 0.7$ in (c) and $\eta = 0.9$ in (d).

The efficiency of the detection $\eta$ is a parameter from the feedback instead of the one from the qubit system; therefore we hold the view that the feedback can compensate the QFI of estimated parameter $\eta$. As shown above, after a period of time, the QFI about $\eta$ almost does not change with time. That is to say, the information of the estimated parameter the system dissipates is exactly equal to the one that the feedback compensates, the system reach a dynamic balance. In addition it can be seen from the Fig.2 that the optimal feedback can be obtained as $-\sigma_y$ in the initial evolution of the system and the deduction has no relationship with the value of $\eta$.

Making $F = -\sigma_y$, choosing $|\varphi_0\rangle = \cos[\alpha]|0\rangle + \sin[\alpha]|1\rangle$ as initial state, and taking advantaging of Eq. (9), one can obtain the density matrix with the elements

$$\rho_{11}(t) = \frac{-2\tau+e^t(\eta-(\eta-2)\cos[2\alpha])}{2\tau(\eta-2)}, \quad (14)$$

$$\rho_{12}(t) = \rho_{12}{}^*(t) = \frac{1}{2\tau}e^{3t/2}\sin[2\alpha], \quad (15)$$

where $\tau = \exp(2t/\eta)$. The analytic expression of QFI can be acquired by substituting Eqs. (14)

and (15) into Eq. (6). But the expression is very complex, and therefore we only demonstrate it in the form of figure. For different values of $\eta$, one can obtain the evolution of the QFI with respect to $\alpha$, which relates to the initial state, and the time t in Fig.3.

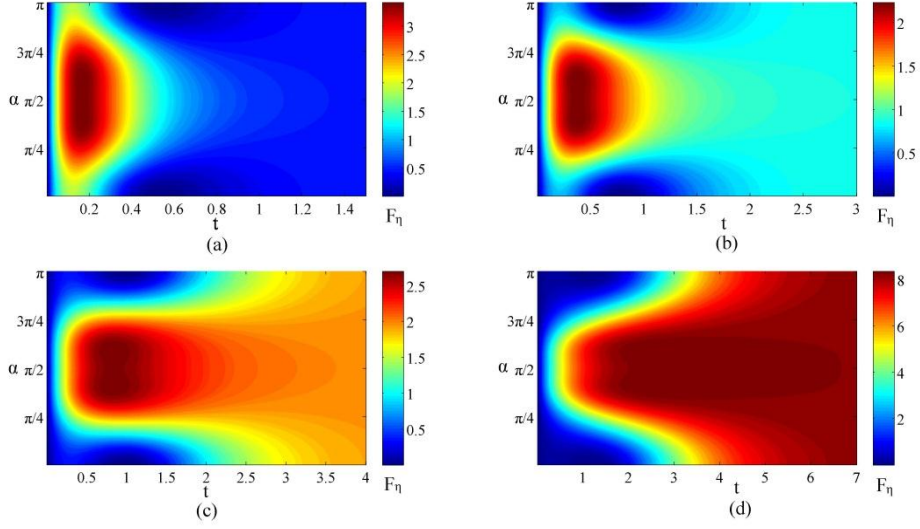

Fig.3: The evolution of QFI against $\alpha$ and t, we take $\eta = 0.3$ in (a), $\eta = 0.5$ in (b), $\eta = 0.7$ in (c), and $\eta = 0.9$ in (d)

The structures of Fig.3 indicate that the bigger the proportion of $|1\rangle$ is, the bigger the QFI is no matter what value the $\eta$ is. The physical interpretations of the above results can be easily given. Due to the spontaneous emission of two level atom, bigger the proportion of $|1\rangle$ is, the more quantum information will be dissipated to the detector thus the bigger the QFI is. In addition, another conclusion, that the system will reach the dynamic balance and the QFI of the dynamic balance has nothing to do with the initial state we choose, can be deduced.

What follows is the discussion on the case that the system reaches the dynamic balance. When the system reaches the dynamic balance, namely $t \to \infty$, the analytical expression of QFI for arbitrary initial state will be acquired:

$$\lim_{t\to\infty} F_\eta = \frac{\lambda^2(1+2\lambda\cos[\beta])^2}{(\eta+\lambda^2+2\eta\lambda\cos[\beta])(\eta+2\lambda^2+2\eta\lambda\cos[\beta])^2} \qquad (16)$$

It can be easily derived from Eq. (16) that the QFI of the dynamic balance has nothing to do with $\alpha$, in other words, the initial state we choose has no effect on the accuracy of estimated parameter $\eta$. According to Eq. (16), the trend of $\lim_{t\to\infty} F_\eta$ with respect to the feedback parameters for different detection efficiency is shown in Fig.4,

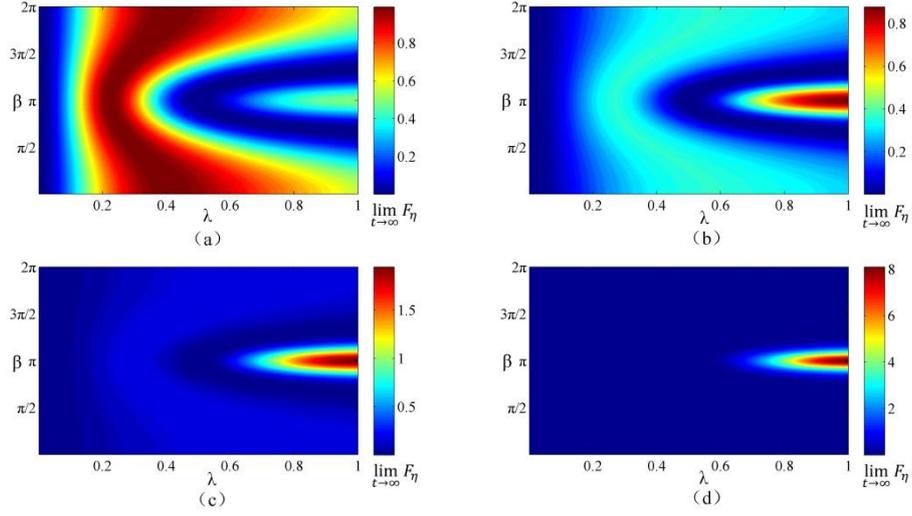

Fig.4: The evolution of the QFI with respect to $\lambda$ and $\beta$ which are related to the feedback, take $\eta = 0.3$ in (a), $\eta = 0.5$ in (b), $\eta = 0.7$ in (c) and $\eta = 0.9$ in (d).

As shown in Fig.4, the QFI can be strengthened by carefully tuning the feedback $\lambda(\sin[\beta]\sigma_x + \cos[\beta]\sigma_y)$ to an optimal value when the system reaches the dynamic balance. On the other hand, one can deduce that the best feedback for a higher detection efficiency ($\eta \geq 0.38$) is obtained as $F = -\sigma_y$, and the advantage of the optimal feedback $F = -\sigma_y$ is more and more obvious with the increase of $\eta$. In addition, taking $\lambda = 1$, the evolution of $\lim_{t\to\infty} F_\eta$ with $\eta$ for feedback $\sin[\beta]\sigma_x + \cos[\beta]\sigma_y$ can be obtained in Fig.5

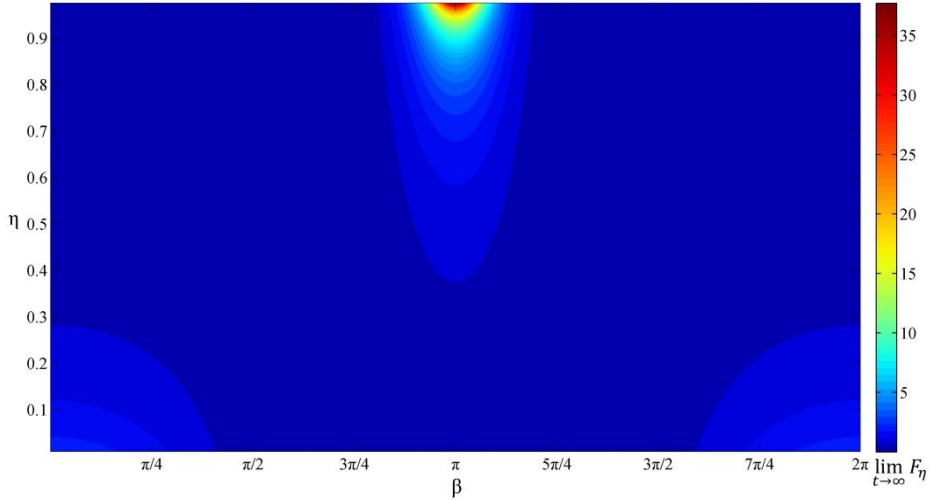

Fig.5: The evolution of QFI in the dynamic balance with $\eta$ and $\beta$.

As shown in Fig.5, the QFI has a positive relation with the value of $\eta$ in the state of dynamic balance when we choose the optimal feedback $F = -\sigma_y$, while the relation becomes negative one when we choose $\sigma_x$ or $\sigma_y$ as feedback.

### C. The feedback $\lambda\sigma_z$

As a comparison, the feedback $\lambda\sigma_z$ will be investigated in the following. Substituting $F = \lambda\sigma_z$ and $|\varphi_0\rangle = \cos[\alpha]|0\rangle + \sin[\alpha]|1\rangle$ into Eq. (9), one can obtain the function of density matrix with the elements

$$\rho_{11}(t) = e^{-t}\sin[\alpha]^2 \qquad (17)$$

$$\rho_{12}(t) = \frac{e^{-t/2}\sin[2\alpha]}{2\Gamma} - i\frac{8e^{-t}(e^{t/2}-\Gamma)\eta\lambda\sin^2[\alpha]}{2\Gamma(\eta-4\lambda^2)} \qquad (18)$$

where $\Gamma = \exp(2t\lambda^2/\eta)$. Choosing $|1\rangle$ as initial state and taking advantaging Eqs. (6), (17), and (18), we can obtain the analytic form of QFI. But expression is very complex, and therefore we only demonstrate it in the form of figure. The time evolution of $F_\eta$ is shown in Fig.6:

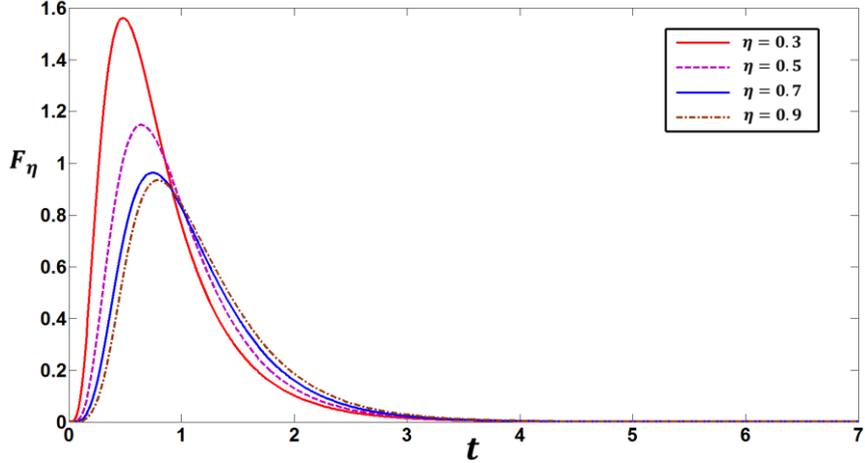

Fig.6: Time evolution of the QFI for the different value of $\eta$, here we take $\lambda = 1$, and choose $|1\rangle$ as initial state.

As shown in Fig.6, unlike the QFI of the feedback $\lambda(\sin[\beta]\sigma_x + \cos([\beta]\sigma_y)$, the one of $\lambda\sigma_z$ tend to zero with evolution of time for the different value of $\eta$. That is to say, the feedback $\sigma_z$ cannot compensate the information of the estimated parameter $\eta$ that the system dissipates. Compared with the Fig.2, the maximum value of QFI in Fig.6 is smaller, and therefore we hold the point that the feedback $\lambda(\sin[\beta]\sigma_x + \cos([\beta]\sigma_y)$ can more effectively enhance the accuracy of the parameters to be estimated than the feedback $\lambda\sigma_z$. In addition, as we know higher the QFI means the better the accuracy, and therefore the maximum of the QFI in the evolution is more meaningful than the QFI at other moment. Thus we denote the maximum value of the QFI in the time evolution by $\max_t F_\eta$, and the evolution of $\max_t F_\eta$ with respect to the parameter of feedback is obtained in Fig.7. The monotone structure implies that the $\max_t F_\eta$ can be strengthened by carefully tuning the feedback strength $F = \lambda \sigma_z$ to an optimal value.

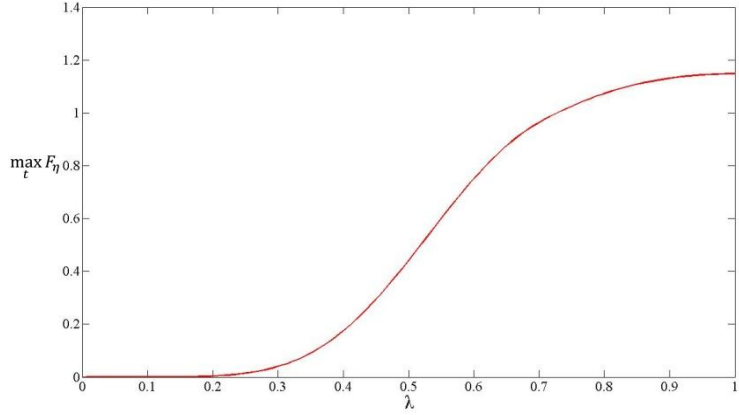

Fig.7: The evolution of $\max_t F_\eta$ with respect to $\lambda$, here we take $\eta = 0.5$.

For different values of $\eta$, one can obtain the evolution of the QFI with respect to the time and the initial state we choose in Fig.8.

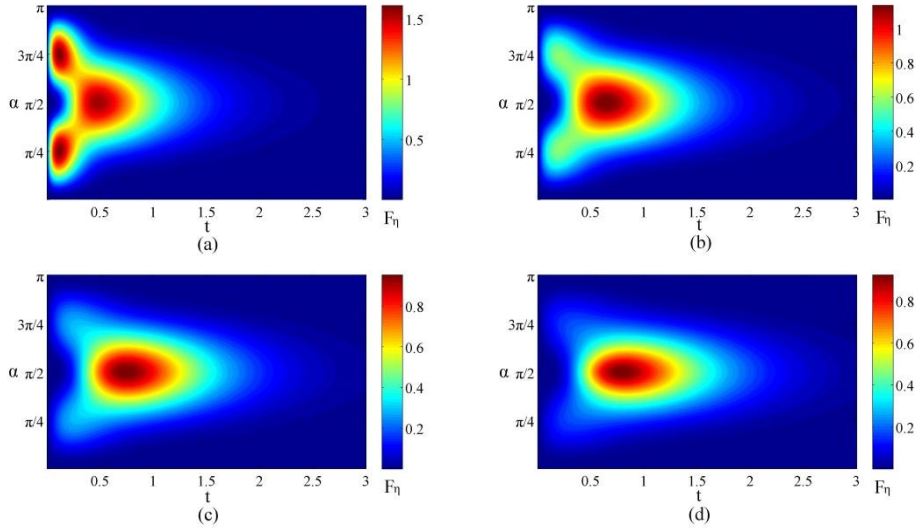

Fig.8: The evolution of QFI against $\alpha$ and t, we take $\eta = 0.3$ in (a), $\eta = 0.5$ in (b), $\eta = 0.7$ in (c), and $\eta = 0.9$ in (d).

It can be seen from Fig.8 that, when the value of $\eta$ is relatively small, the QFI will reach the maximum value at the case that we choose $(|0\rangle \pm |1\rangle)/\sqrt{2}$ as initial state. However the optimal initial state becomes $|1\rangle$ with the increase of $\eta$. In addition, we have the QFI is equal to zero all the time when $|0\rangle$ is chose as initial state that is to say the information of the $\eta$ will not be acquired if we choose $|0\rangle$ as initial state.

## V. CONCLUSIONS

In conclusion, we have investigated the effects of three different constant feedback operator types on precision of the detection efficiency. First of all, we find that the feedback $F = -\sigma_y$ can enhance the accuracy of the parameters to be estimated more effectively than other feedback types, and the bigger the proportion of $|1\rangle$ is the bigger the QFI is no matter what value the $\eta$ is in the initial evolution. On the other hand, when the system reach the dynamic balance which has nothing to do with the initial state, one can find that that $F = -\sigma_y$ is the optimal feedback only for higher detection efficiency and the advantage of the optimal feedback is more and more obvious with the increase of $\eta$. In addition, the QFI of dynamic balance has a positive relation with detection efficiency for the optimal feedback $F = -\sigma_y$ while the relation becomes negative one when we choose $\sigma_x$ or $\sigma_y$ as feedback. At last, when we choose the feedback $\lambda \sigma_z$ the maximum value of QFI has a positive relation with the feedback strength and the information of the parameter to be estimated will not be acquired if we choose $|0\rangle$ as initial state.

**Acknowledgments**

This work is supported by the National Natural Science Foundation of China (Grant No. 11574022).

**Appendix:**

When homodyne detection are performed on the system, the stochastic evolution of the system is described as [11,28-32]:

$$d\rho_c(t) = \{dN_c(t)\mathcal{G}[c + \varpi] + dt\mathcal{H}[-iH - \beta c - \frac{1}{2}c^\dagger c]\}\rho_c(t) \tag{A1}$$

in which $c$ is the annihilation operator of the cavity, the point process $dN_c(t)$ is the increment (either zero or one) in the photon count $N_c(t)$ in the time interval$(t, t + dt)$, $\rho_c(t)$ stands for the quantum state conditioned on the measurement, and $\varpi$ is a complex number representing the coherent amplitude. Here the superoperators $\mathcal{G}$ and $\mathcal{H}$ are defined as[21]:

$$\mathcal{G}[R]\rho = \frac{R\rho R^\dagger}{\text{Tr}[R\rho R^\dagger]} - \rho \tag{A2}$$

$$\mathcal{H}[R]\rho = R\rho + \rho R^\dagger - Tr[R\rho + \rho R^\dagger]\rho \tag{A3}$$

where $R$ represents an arbitrary operator. Let $\varpi$ be real so that the homodyne detection leads to a measurement of the x quadrature of the system dipole. The rate of photodetections at the detector is $E[dN_c(t)] = \text{Tr}[(\varpi^2 + \varpi x + c^+c)\rho_c(t)]dt$ with $x = c + c^\dagger$. The $dN_c(t)$ is a point process in Eq.(A1), it is possible to approximate the photocurrent by a continuous function of time in the limit that the local oscillator amplitude goes to infinity, and then the smooth evolution is governed by the stochastic master equation[33]:

$$dρ_c(t) = -i[H,ρ_c]dt + D(c)ρ_c dt + dW(t)\mathcal{H}(c)ρ_c \qquad (A4)$$

where $dW(t)$ is the standard Wiener increment with mean zero and variance $dt$. As we know, the dynamics of a controlled system is based on the information obtained from the system through measurement. In the following we will focus on the continuous feedback control, and take the Markovian feedback of the white-noise measurement record via a Hamiltonian. The continuous measurement record can be described by the homodyne detection photocurrent [29 30.32]

$$I(t) = \text{Tr}[xρ_c(t)] + ξ(t) \qquad (A5)$$

in which $ξ(t)=dW/dt$ is real $δ$-correlated noise. Then the control Hermitian can be written as $H_{fb} = I(t)F$ with $F = F^†$ is the feedback operator. The stochastic equation for the conditioned system state including feedback is:

$$dρ_c(t) = dt\{-i[H,ρ_c] + D(c)ρ_c - i[F,cρ_c + ρ_c c^†] + D[F]ρ_c\} + dW(t)\mathcal{H}[c - iF]ρ_c \qquad (A6)$$

The last term can be omitted when we take the ensemble average of the master equation. Therefore we obtain:

$$\frac{dρ}{dt} = -i\left[H + \frac{1}{2}(c^†F + Fc), ρ\right] + D(c - iF)ρ \qquad (A7)$$

where $ρ$ is the ensemble average of $c$. The effect of the feedback is thus seen to replace $c$ by $c - iF$, and to add an extra term to the Hamiltonian.